\documentclass{ieeeaccess}

\usepackage[utf8]{inputenc}
\usepackage{cite}
\usepackage{amsmath,amssymb,amsfonts}
\usepackage{algorithmic}
\usepackage{graphicx}
\usepackage{textcomp}
\usepackage{comment}

\usepackage{listings}
\usepackage{siunitx}
\usepackage{color}
\usepackage{enumitem}
\usepackage{booktabs}
\usepackage{gensymb}

\newcommand{\PaperTitle}{Hybrid Boolean Networks as Physically Unclonable Functions}

\renewcommand{\thesubsection}{\thesection.\Alph{subsection}}

\DeclareMathOperator*{\argmax}{arg\,max} 

\newcommand{\Nchips}{8}

\newcommand{\Nrepeats}{100}

\definecolor{codegreen}{rgb}{0,0.6,0}
\definecolor{codegray}{rgb}{0.5,0.5,0.5}
\definecolor{codepurple}{rgb}{0.58,0,0.82}
\definecolor{backcolour}{rgb}{0.95,0.95,0.92}
\lstdefinestyle{mystyle}{
    backgroundcolor=\color{backcolour},   
    commentstyle=\color{codegreen},
    keywordstyle=\color{magenta},
    numberstyle=\tiny\color{codegray},
    stringstyle=\color{codepurple},
    basicstyle=\footnotesize,
    breakatwhitespace=false,         
    breaklines=true,                 
    captionpos=b,                    
    keepspaces=true,                 
    numbers=left,                    
    numbersep=5pt,                  
    showspaces=false,                
    showstringspaces=false,
    showtabs=false,                  
    tabsize=2
}
\def\BibTeX{{\rm B\kern-.05em{\sc i\kern-.025em b}\kern-.08em
    T\kern-.1667em\lower.7ex\hbox{E}\kern-.125emX}}

\usepackage{hyperref}

\begin{document}

\history{Date of publication xxxx 00, 0000, date of current version xxxx 00, 0000.}

\doi{10.1109/ACCESS.2017.DOI}

\title{\PaperTitle}

\author{\uppercase{Noeloikeau Charlot}\authorrefmark{1},
\uppercase{Daniel Canaday\authorrefmark{2}, Andrew Pomerance\authorrefmark{3}, and Daniel J. Gauthier},\authorrefmark{4} (Member, IEEE)}

\address[1,4]{Ohio State University, Department of Physics, 191 West Woodruff Ave, Columbus, OH 43202, USA, (e-mail: charlot.5@osu.edu, gauthier.51@osu.edu)}

\address[2,3]{Potomac Research, LLC, 801 N Pitt St \#117, Alexandria, VA 22314, USA (e-mail: daniel@potomacresear.ch, andrew@potomacresear.ch)}

\tfootnote{This work was supported by the Department of the Army through award number W31P4Q-19-C-0014 and by (for NC and DJG) Asymmetric Technologies, LLC through the project `Resilient and enhanced security UAS flight control' supported by the Ohio Federal Research Network. A patent was filed based on this work under PCT/US2020/027072 for Systems and methods using hybrid Boolean networks as physically unclonable functions.}

\markboth
{Noeloikeau Charlot \headeretal: \PaperTitle}
{Noeloikeau Charlot \headeretal: \PaperTitle}

\corresp{Corresponding author: Noeloikeau Charlot (e-mail: charlot.5@osu.edu).}

\begin{abstract}
We introduce a Physically Unclonable Function (PUF) based on an ultra-fast chaotic network known as a Hybrid Boolean Network (HBN) implemented on a field programmable gate array. The network, consisting of $N$ coupled asynchronous logic gates displaying dynamics on the sub-nanosecond time scale, acts as a `digital fingerprint' by amplifying small manufacturing variations during a period of transient chaos. In contrast to other PUF designs, we use both $N$-bits per challenge \textit{and} obtain $N$-bits per response by considering challenges to be initial states of the $N$-node network and responses to be states captured during the subsequent chaotic transient. We find that the presence of chaos  amplifies the frozen-in randomness due to manufacturing differences and that the extractable entropy is approximately $50\%$ of the maximum of $N2^{N}$ bits.  We obtain PUF uniqueness and reliability metrics $\mu_{inter}$ = 0.40$\pm$0.01 and $\mu_{intra}$ = 0.05$\pm$0.00, respectively, for  an $N=256$ network. These metrics correspond to an expected Hamming distance of 102.4 bits \textit{per response}.  Moreover, a simple cherry-picking scheme that discards noisy bits yields $\mu_{intra} < 0.01$ while still retaining $\sim200$ bits/response (corresponding to a Hamming distance of $\sim80$ bits/response).  In addition to characterizing the uniqueness and reliability, we demonstrate super-exponential scaling in the entropy up to $N=512$ and demonstrate that PUFmeter, a recent PUF analysis tool, is unable to model our PUF.  Finally, we characterize the temperature variation of the HBN-PUF and propose future improvements.
\end{abstract}


\titlepgskip=-15pt

\maketitle

\section{Introduction}
\label{sec:introduction}
\PARstart{P}{hysically} unclonable functions (PUFs) are an emerging technology that extract randomness, or entropy, from uncontrollable manufacturing variations in the physical structure of identically produced devices \cite{PUFDEF,TAXONOMY}. PUFs use this entropy to reliably generate a `digital fingerprint' - a unique sequence of 0's and 1's known as a bitstream - that is produced by the device but never stored \cite{CRYPTOKEY}. In practice, PUFs are often circuits embedded in other devices that reliably map an input (or \textit{challenge}) to an output (or \textit{response}) in a way that is unique to a particular copy (or \textit{instance}) of the device. 

For example, the start-up behavior of static random-access memory (SRAM) produces an identifying bit pattern suitable for use as a PUF \cite{SRAM}. Ideally, this identifying behavior cannot be reproduced (or cloned), either because it is physically impossible to recreate the same conditions in another device, or because it is mathematically impossible to accurately predict the PUF's behavior. In summary, we highlight three practical properties of PUFs:

\begin{itemize}
    \item \textbf{Uniqueness}: Responses from different instances to the same challenge are different enough to distinctly identify each instance;
    \item \textbf{Reliability}: Responses from an individual instance to the same challenge are similar enough to consistently identify that instance;
    \item \textbf{Unclonability}: The challenge-response pairs (CRPs) of an individual instance cannot be: (1) physically replicated by another instance, or (2) inferred from knowledge of the device manufacturing process or previously revealed CRPs.
\end{itemize}

In early work, PUFs were constructed using complex optical scattering devices or custom fabricated silicon chips \cite{EARLYPUF}. More recently, there is an industry trend toward using reprogrammable devices such as field-programmable gate arrays (FPGAs) for PUF-based IP protection. For example, IntrinsicID offers the commercially available `butterfly PUF,' an SRAM-PUF embedded directly into some manufacturers higher-end FPGAs \cite{FPGAPUF}. 

However, SRAM-PUFs, such as the butterfly PUF, are `weak' in the sense that there are relatively few CRPs obtainable per device (in this case resulting from the static initialization of each memory cell at power-up) \cite{TAXONOMY}. As a result, their use for authentication purposes are limited because an attacker can clone the device by obtaining the full set of CRPs in a short amount of time. `Strong' PUFs, on the other hand, contain a relatively large number of independent CRPs, making attempts to extract or predict all of them a difficult or impossible task \cite{STRONGPUF}. Moreover, the design and practical implementation of strong FPGA-based PUFs remains an open problem \cite{PUFDEF}.

Modern PUF proposals have also started to explore chaotic dynamics as an additional source of entropy \cite{CHAOSPUF,CHAOSFPGA,CHAOSENHANCE}. Chaos is characterized by a exponential divergence between initially similar trajectories. As discussed in more detail below, this behavior can be used by a PUF to amplify the entropy available from the small physical variations inherent in any manufacturing process. Moreover, we hypothesize that chaos provides resilience to machine learning due to the existence of `fractal basin boundaries' \cite{EDSBOOK}, which is a phenomenon in chaotic systems in which dividing lines between different behaviors have a fractal structure. In the standard interpretation, this means that an infinitesimal change in the initial conditions of the system does not yield a smooth change in the asymptotic behavior of the system; instead, the system may evolve to a disjoint attracting set.  We hypothesize that a chaotic PUF has a similar behavior with respect to the system parameters, such that an infinitesimal change does not yield a smooth change in the measured response.  Hence, even marginal uncertainty in the system parameters changes the entire class of possible outcomes, likely confusing attempts at prediction.

Finally, many PUFs incorporate asynchronous (unclocked and analog-like) logic into their design \cite{TAXONOMY,PUFDEF}. Asynchronous logic can require fewer resources (time, area and power) than conventional synchronous circuits governed by a global clock.   Morevoer, compared to synchronous designs, asynchronous designs are much more sensitive to manufacturing variations.  This is because clocked operations are stabilized by waiting an entire clock period before the next operation, so that any variations in, \textit{e.g.}, rise time or signal propagation time are eliminated.  On the other hand, dynamical properties of even simple unclocked systems such as the frequency of a ring oscillator depend sensitively on variations in rise and fall times.  In general, combinatorial loops can be designed that operate at the maximum frequency allowed by the hardware, where the dynamics are most sensitive to manufacturing variations. Thus, asynchronous PUFs are useful as compact, low-power cryptography primitives.

\subsection{This Work}
\label{subsec:PAPER}

In this paper, we propose a design for a strong, chaos-enhanced, asynchronous PUF and demonstrate its implementation on an FPGA. Our PUF is based on a network of coupled, unclocked logic gates known as an autonomous Boolean network (ABN) combined with a clocked digital control and readout layer, forming what we call a hybrid Boolean network (HBN, HBN-PUF). The HBN-PUF can be incorporated into existing FPGA designs without specialized hardware, having a resource count proportional to the number of nodes in the network $N$. The unique properties of the HBN-PUF compared to existing strong PUF proposals are:

\begin{itemize}
    \item  The HBN-PUF produces $N$ (or potentially more) response bits per $N$-bit per challenge.   Thus, extracting secrets of a given length requires $\sim 1/N$ the number of queries, which translates into time, storage and network traffic efficiency.  Moreover, the additional bits per response can be used for error correction and improving environmental resilience, and the multi-dimensional response space and possible fractal basin boundaries will likely frustrate machine learning attacks.  
    \item Unlike many conventional PUFs, such as delay-line PUFs \cite{TAXONOMY}, the HBN-PUF does not require carefully constructed circuit paths with specified delay characteristics; rather, automatic placement of circuit elements by standard vendor-supplied compilation and synthesis tools yield usable HBN-PUFs.
    \item The ABN part of the HBN-PUF exhibits picosecond-scale asynchronous transient-chaotic dynamics.  Because of these ultra-fast dynamics, response readout occurs in less than 10 ns, which has important practical applications because the number of CRPs required for strong industrial-scale enrollment can be obtained in a short time.

\end{itemize}

The paper is organized as follows. Our proposed HBN-PUF design is described in Sec. \ref{sec:ABNPUF}, with a discussion of the circuit and data collection process in Sec. \ref{subsec:DESIGN} and the physical origins of PUF behavior in Sec. \ref{subsec:BEHAVIOR}. Section \ref{sec:STATS} is devoted to experimentally characterizing the HBN-PUF behavior by measuring its uniqueness and reliability (\ref{subsec:MU}), entropy scaling (\ref{subsec:ENTROPY}), resilience to machine learning (\ref{subsec:AI}), and temperature variation (\ref{subsec:TEMP}). Section \ref{sec:CONC} concludes with a brief discussion and future work. Supporting materials are given in the Appendix, including the hardware description language code that instantiates our design.

\section{Proposed HBN-PUF Design}
\label{sec:ABNPUF}
\Figure[!h](topskip=0pt, botskip=0pt, midskip=0pt)[width=5.in]{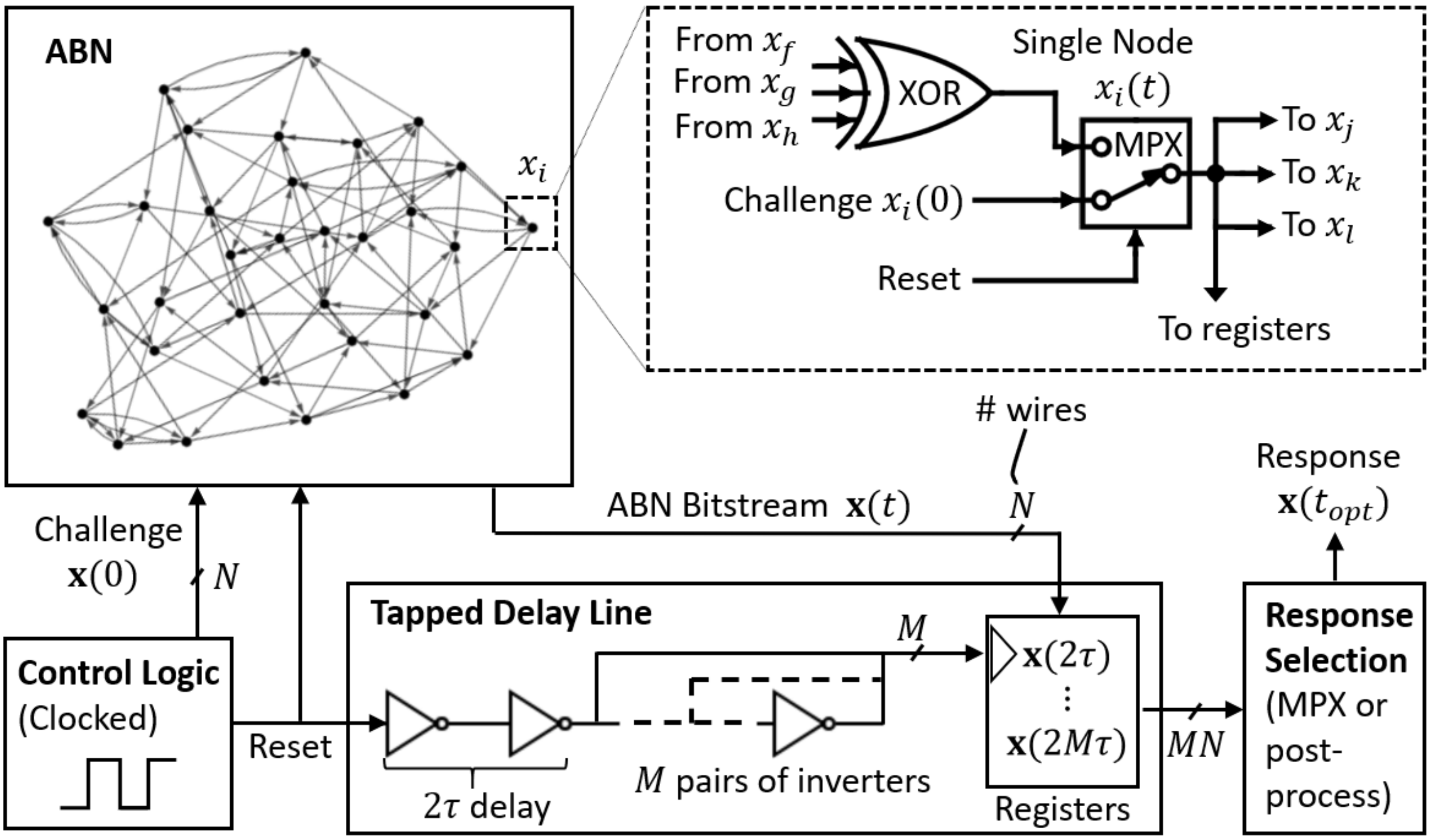}
{Proposed HBN-PUF design. The studied XOR-ABN topology is shown in the upper left, and the logic of an individual node in the upper right. Shown in the bottom are the clocked logic used to apply the challenge and the tapped delay line used to select the response. The specific connections between nodes (i.e, the identity of nodes $f-l$) are governed by the topology of the network; shown is an $N=32$ regular random graph of degree 3.\label{fig:HBNPUF}}

Our proposed design is shown in Fig. \ref{fig:HBNPUF}, which consists of a network of $N$ coupled `nodes' and a clocked digital readout and control layer, forming an HBN. Each node is a combinatorial logic circuit that takes as input the outputs of other nodes in the network and a global reset signal; we refer to the output of this circuit as the `state' of each node in the network.  When the reset signal is high, the state of the node is the corresponding bit in the challenge string, and the state of the entire network is exactly equal to the challenge string $\mathbf{C}$.  When the reset signal is low, the state of the node is given by the XOR of the states of its input nodes and the entire network is a large recurrent combinatorial loop that evolves in time without a clock (\textit{i.e.}, autonomously).  Typical clocked digital logic circuits constrain voltages to be near logic high or low most of the time so that the output voltages of a gate is near logic high or low.  In contrast, the individual semiconductor devices in an ABN act as highly nonlinear input-output devices with analog (but Boolean-like) dynamics, and the voltages take on a continuous range of values between logic-high and -low.  During this time, the digital readout layer captures a Booleanized representation of the true analog network state in discrete time intervals. A single state at the optimal time of measurement is then selected as the response $\mathbf{R}$. 

In contrast to other PUF designs, we stress that the challenge and response are both $N$-bit strings, specifying the network's initial condition and Booleanized state in a chaotic transient, respectively. Thus, there are $N$ response bits for each of $2^{N}$ challenges. Hence, the number of extractable bits from the HBN-PUF may scale super-exponentially as $N2^{N}$, yielding a strong PUF.

\subsection{Design Specifics and Data Collection}
\label{subsec:DESIGN}
For the specific HBN considered in this work, each node takes exactly 3 inputs, and the combinatorial function is the 3-input XOR, as shown in the upper right of Fig. \ref{fig:HBNPUF}. Both of these design choices are flexible.  The XOR function is chosen because it is maximally sensitive to its inputs, and the output is balanced between high and low; the overall bias of the response can be controlled by replacing the XOR with a Boolean function that has more or fewer high outputs.  Three inputs were chosen in order to fit within a Cyclone V logic element; more or fewer inputs can be used to match the layout to other FPGA architecture details.  Moreover, structure can be applied to the network (such as ring topologies \cite{HBNTRNG}) to fine tune statistical and performance properties of the resulting response.  These aspects will be explored in follow up papers, but in this work each node's XOR gate takes the output of three nodes ($f,g,h$ in Fig. \ref{fig:HBNPUF}), randomly chosen without replacement from among the $N-1$ other nodes, and in turn its multiplexer feeds the XOR gate of three other nodes ($j,k,l$ in Fig. \ref{fig:HBNPUF}).  When the clocked reset signal is low, the multiplexer passes the node's XOR gate. When the reset signal is high, the node's multiplexer holds the initial condition, which is given by a corresponding bit of the challenge. In this way, the analog state of all nodes in the network $\textbf{x}(t)=\{x_{i}(t)\}_{i=1}^{N}\in[0,1]^{N}$ are initially held fixed to the digital $N$-bit challenge string $\textbf{C}$, described mathematically as
\begin{equation}
    \label{CHALLENGE}
    \textbf{x}(0)=\textbf{C} \in\{0,1\}^{N}.
\end{equation}

The HBN stabilizes to the initial condition nearly instantaneously, but we hold it there for several $\sim100$ MHz clock cycles of holding Reset high. The dynamics are then enabled by setting the Reset signal low, causing each multiplexer to pass the output of the autonomous XOR gate that feeds it. The network then evolves continuously in time and each XOR gate updates asynchronously based on the analog voltage of its neighbors. 

During this time, the HBN dynamics are measured by sending the Reset signal down $M$ pairs of inverter gates (\textit{i.e.}, a delay line). An associated register is triggered after the delayed Reset signal passes over a given pair of inverters. Each register Booleanizes the analog state of the HBN at that time and stores it digitally. This results in a sequence of $N$-bit Boolean state vectors in memory recording the bitstream produced by the network $\{\textbf{x}(2\tau), \textbf{x}(4\tau), ..., \textbf{x}(2M\tau)\}\in\{0,1\}^{NM}$. 

Here, $\tau\sim0.25$ ns is the mean delay time of a single inverter-gate, which is similar to the timescale of the XOR gate and multiplexer operations. Thus, the bitstream is sampled at a similar rate as the HBN dynamics, in roughly $2\tau\sim0.5$ ns intervals. However, like all logic elements, each delay is subject to manufacturing variation, and so the sampling rate is not completely uniform. This also contributes to the manufacturing variation that gives rise to PUF behavior.  Moreover, by using pairs of inverter gates rather than a clock source, the delay through the delay line varies with temperature and voltage in a similar way to the dynamical timescale of the nodes in the network.  Thus, the delay line imparts some robustness to environmental variation.

The response $\textbf{R}$ is selected from among this bitstream as a single state of the network at an optimal point in time $\textbf{x}(t_{opt})$ during the chaotic transient
\begin{equation}
    \label{RESPONSE}
    \textbf{R} = \theta(\textbf{x}(t_{opt})) \in\{0,1\}^{N},
\end{equation}
where $\theta: [0,1]^{N}\rightarrow\{0,1\}^{N}$ is an element-wise thresholding operation, corresponding to the Booleanization of the real-valued $\textbf{x}(t)$ performed by the registers.  The details of determining $t_{opt}$ are discussed in Sec. \ref{sec:STATS}.

\subsection{HBN Dynamics and PUF Behavior}
\label{subsec:BEHAVIOR}
If each logic gate in an HBN were synchronously updated by a global clock, it would execute the digital Boolean XOR function exactly, and node states would take on discrete values 0 or 1 at each clock cycle.  In this mode, the state at each discrete time step would be exactly determined by the $N$-bit Boolean state at the previous time step, and the entire network would act as a pseudo-random number generator.  However, because the logic gates are unclocked, their inputs can change at the same time that they are transitioning between logic high and low. As a result, nodes have the potential to take on intermediate logic values (analog voltages) \cite{ROSINTHESIS}.  Thus, the dynamics of nodes are better described by continuous differential equations that model the rise and fall times resulting from the finite capacitances and resistances in the devices, and not by discrete Boolean dynamics.  Moreover, the state at a specific time is not given by the states of its inputs at the current time, but rather by time-delayed versions, due to the finite speed at which signals propagate along interconnects.  Taken together, this causes the asynchronous XOR gate to behave as a highly nonlinear input-output device that multiplies signal edges, which quickly causes the dynamics to reach the maximum switching frequency allowed by the hardware \cite{BCHAOS1, GHIL}. 

Under these conditions, the network dynamics become highly sensitive to amplitude fluctuations about the intermediate voltage value. Here, small perturbations to the voltage at the XOR gate, such as those due to manufacturing variation, noise, and differences in initial conditions, will cause the time at which the node switches between logic high and low to vary, resulting in previously similar waveforms diverging. As a result, ABNs consisting of XOR gates can exhibit chaos even in small networks\cite{BCHAOS2}. When combined with a digital readout and control layer to form an HBN, they have been used as ultra-fast true-random-number generators (TRNGs) capable of a $12.8$ Gbit/s entropy rate\cite{HBNTRNG}.

Based on past research and the discussions above, we identify three sources of entropy in XOR-HBNs related to PUF behavior:
\begin{enumerate}
    \item Frozen-in heterogeneity (manufacturing differences),
    \item Thermal and charge fluctuations (noise), and
    \item Deterministic chaos (unpredictability and nonlinear amplification of timing differences)
\end{enumerate}
Each source of entropy produces variations in the bitstream generated by the digital readout layer of the clocked portion of the network. However, each source has a separate physical origin as discussed in the rest of this section.

\Figure[!htb](topskip=0pt, botskip=0pt, midskip=0pt)[width=5.0in]{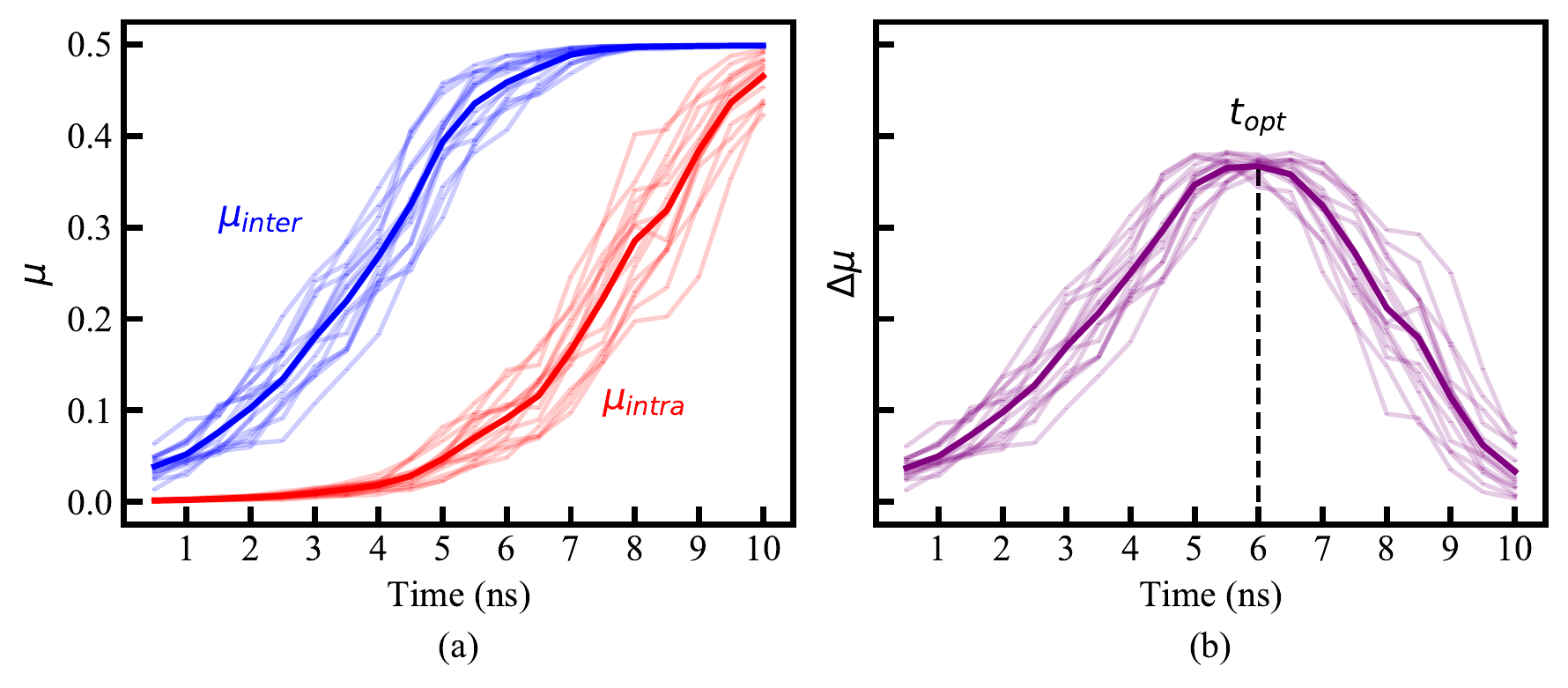}
{(a). $\mu_{inter}$ (blue) and $\mu_{intra}$ (red) vs time.  Light lines correspond to the metrics calculated on a single PUF class (\textit{i.e.}, choice of random network and placement on an FPGA) for an $N=256$ node HBN-PUF, and the dark lines correspond to the average values calculated over all PUF classes (\textit{i.e.}, the average expected behavior for an $N=256$ random, 3-input HBN-PUF). (b) $\Delta\mu_{intra}$ vs time, with same definitions for light and dark lines.  The highlighted $t_{opt}$ is the time at which the average $\Delta \mu$ is maximum.  We note, however, that there is some variation in the specific $t_{opt}$ for each PUF class.\label{fig:MUvsT}}

Frozen-in heterogeneity is due to small variations in the physical properties of the wiring and logic elements and it is this source of entropy that forms the primary basis of PUF behavior. Slight physical differences between nodes and wires - such as node input impedence, switching rate, and signal propagation time - alter the time at which the analog voltage of individual nodes cross the logic threshold for nominally identical inputs. The effect of these manufacturing variations are more pronounced at the ultra-fast time scale of the dynamics, which become distinctly correlated with the unique physical characteristics of an individual device. Such correlations produce the identifying information used to distinguish different FPGAs programmed with the same HBN design. They are quantified by the uniqueness parameter $\mu_{inter}$ (Appendix \ref{app:STATS}, \eqref{intereq}).

Thermal and charge fluctuations are sources of time-dependent stochastic behavior (often referred to as `noise'), which reduce the reliability of the PUF.  Noise perturbs the amplitude of the logic gates in the asynchronous portion of the network and changes the times at which nodes cross the threshold separating logic high from logic low.   If a transition is near the time at which the readout logic registers the node state, small variations in the threshold crossing time can change a registered zero to a one or vice versa. This alters the bitstream of a single device under repeated measurement, introducing unreliability quantified by $\mu_{intra}$ (Appendix \ref{app:STATS}, \eqref{intraeq}).

Chaotic systems have a positive entropy rate separate from noise and manufacturing variations, which serves to amplify both of these sources of entropy. The entropy attributed to chaos is due to the finite precision of physical measurements and the exponential sensitivity of chaotic systems to initial conditions. Any physical measurement of initial conditions has a necessarily limited precision, and so two trajectories measured to have the same initial conditions will diverge due to the unmeasurable differences in the true initial state of each system. Chaos thereby magnifies any small differences in the applied challenge over time, acting as a nonlinear amplifier of the other sources of entropy and contributing to the unclonability property.  

These three sources of entropy are visible in Fig. \ref{fig:MUvsT}(a), which is a plot of $\mu_{inter}$ and $\mu_{intra}$ vs. measurement time.  Frozen-in heterogeneity is illustrated by the separation between $\mu_{inter}$ and $\mu_{intra}$ at very short measurement times, noise is illustrated by the fact that $\mu_{intra}$ is non-zero, and the effect of chaos is illustrated by the fact that both measures grow exponentially until saturating at 0.5.  In the next section, we discuss finding $t_{opt}$ that balances these competing effects.

\section{ABN-PUF Performance Statistics}
\label{sec:STATS}
To be an effective PUF, the entropy rate due to the frozen-in heterogeneity of the HBN must be greater than the noise-induced entropy rate. This is captured by the metric 
\begin{equation}
    \Delta\mu(t):=\mu_{inter}(t)-\mu_{intra}(t),
\end{equation}
which is plotted vs. time for $N=256$ in Fig. \ref{fig:MUvsT}(b).  There is an optimal time of measurement $t_{opt}$ for which the network has coupled sufficiently to manufacturing variations to act as a unique identifier ($\mu_{inter}\sim1/2$), while remaining unperturbed enough by noise to be reliable ($\mu_{intra}\sim0)$, defined by
\begin{equation}
    t_{opt}:=\argmax_{{t\in[2\tau,2M\tau]}}\Delta \mu (t).
\end{equation}
All future statistics are calculated from the network state at this time. In practice, we find $t_{opt}\sim 2-8$ ns for the networks studied, with slowly increasing $t_{opt}$ with network size $N$. Note that $t_{opt}$ is calculated exactly once over an entire PUF class and represents a characteristic timescale of the HBN dynamics. Further, we do not observe significant variation in $t_{opt}$ or $\Delta \mu$ due to differences in the layout of the network or delay line, as demonstrated in Fig. \ref{fig:MUvsT} and described in Appendix \ref{app:EXP}.

In the remainder of this section, we study the performance statistics of the proposed HBN-PUF, including its reliability and uniqueness (\ref{subsec:MU}), entropy (\ref{subsec:ENTROPY}), resilience to machine learning (\ref{subsec:AI}), and temperature variation (\ref{subsec:TEMP}). Corresponding definitions and experimental procedures are elaborated in Appendices \ref{app:EXP}-\ref{app:TEMP}.

\subsection{Reliability and Uniqueness}
\label{subsec:MU}
Reliability and uniqueness are standard means of gauging PUF performance \cite{PUFDEF}. The average fraction of dissimilar bits between responses of different PUFs to a given challenge is ideally 0.5 (random). It is known as `uniqueness'  and described by $\mu_{inter}$. Likewise, the average fraction of dissimilar bits between responses of a fixed PUF to a given challenge, known as `reliability' ($\mu_{intra}$), is ideally 0 (no error). To gauge these measures, we study the pairwise difference between HBN-PUF responses to various challenges; see Appendix \ref{app:STATS} for details. 

\Figure[!hbt](topskip=0pt, botskip=0pt, midskip=0pt)[width=6.0in]{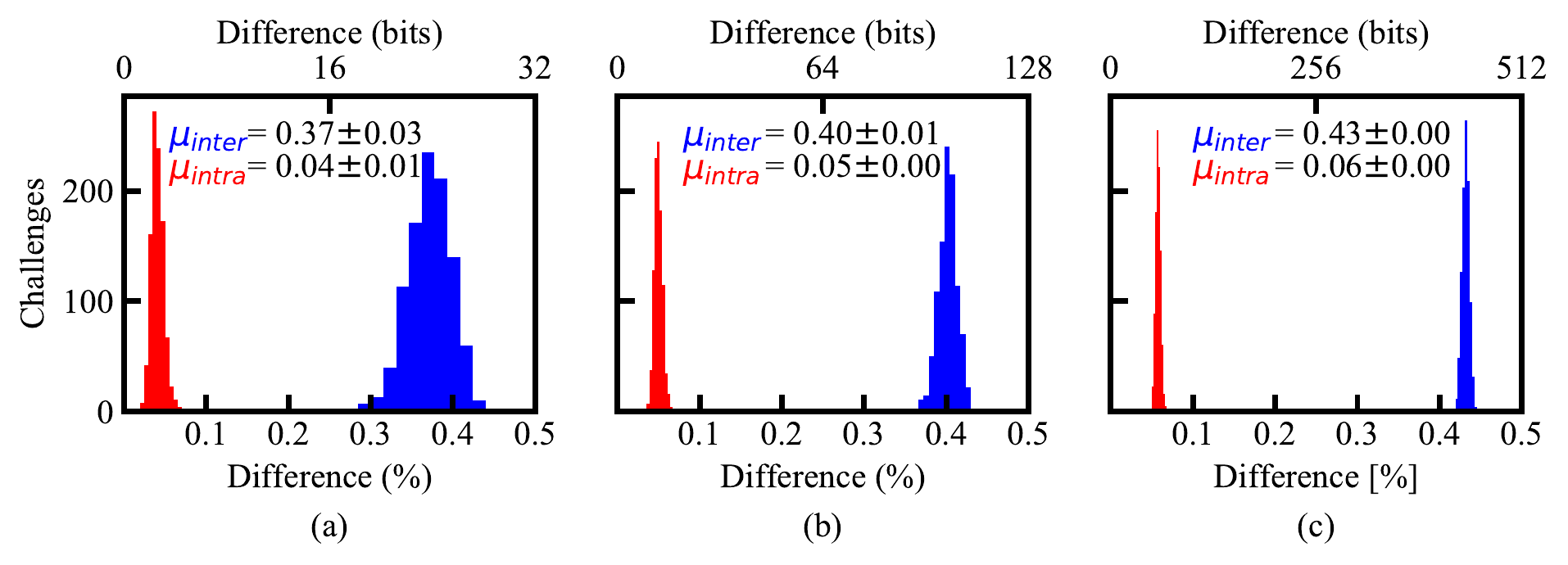}
{Challenge-response histograms for network sizes (a) $N=64$, (b) $N=256$, (c) $N=1024$. We plot \eqref{intraeq} and \eqref{intereq} (see Appendix \ref{app:STATS}), the means of which are $\mu_{intra}$ and $\mu_{inter}$, respectively.\label{fig:MU}}

Figure \ref{fig:MU} shows the number of unique challenge bitstrings yielding response pairs differing on average by the given fraction (bottom axis) or number of bits (top axis) for three different PUF sizes, $N=64$, 256, and 1024. Two histograms are plotted, where the differences are calculated with respect to the same chip (red) and with respect to other chips (blue). Eight different chips were used to estimate $\mu_{inter}$.  There appears clear separation of intra- and inter-device distributions, indicating vanishing false-positive rate for authentication using both network sizes, especially as $N$ increases. This means that our PUF is well-suited to authentication. Furthermore, fewer challenges ($\sim 1/N$) are required for authentication than with single-bit PUFs since the HBN-PUF produces $N$-bit responses.

In practice, we find that $\mu_{intra}$ is driven by a relatively small, fixed subset of nodes (where the subset depends on the chip and the response).  We hypothesize that these nodes are in a metastable state at the measurement time $t_{opt}$, and that a cherry picking error correction scheme \cite{CHERRY} that removes these error-prone bits from the response can be highly effective.  This is illustrated in \ref{subsec:TEMP} and will be studied more extensively in future work.  

\subsection{Exponential Scaling of Entropy with Network Size}
\label{subsec:ENTROPY}
Entropy is of central importance in determining the cryptographic and security properties of a PUF \cite{ENTROPYDEF}. The HBN-PUF, with its multiple bits per response, presents unique challenges to entropy estimation that will be discussed in future work, but in this section we apply previously reported entropy estimation techniques adapted to the HBN-PUF. A PUF can be idealized as a table that gives the response corresponding to a given challenge (called the `CRP table' below). For most strong PUFs, the number of challenges (\textit{i.e.}, the number of rows in the CRP table) grows as $2^N$, and each response is a single bit so the CRP table for a given PUF realization can be described by a binary string of length $2^N$. For the HBN-PUF, on the other hand, each row in the CRP table is itself an $N$-bit string so the entire CRP table is described by an $N2^N$-bit string. Estimating the distribution of binary strings of length $N2^N$ is infeasible even for relatively small $N$; however, we can apply entropy estimates from the PUF literature that make assumptions about this distribution--$H_{min}$, $H_{joint}$, and $H_{CTW}$ (see Appendices \ref{app:MINENT}-\ref{app:CTWENT}). We do not report the values of $H_{CTW}$ below because in nearly all cases it produces full entropy and is never below $H_{min}$ or $H_{joint}$.

The most basic measure is the minimum entropy $H_{min}$, which assumes no correlations between bits and responses and serves as a median. The joint entropy $H_{joint}$ does not assume independence, but does assume that all correlations are pairwise and that no other higher-order correlations exist. Finally, the context-tree weighted entropy $H_{CTW}$ serves as an upper bound by generating a minimum-length compressed binary string encoding the CRP behavior. We plot the first two of these quantities as a function of $N$ in Fig. \ref{fig:ENTROPY} and Table \ref{tab:ENTROPY}, observing that $H_{joint}\leq H_{min}$, which is true by definition.

\Figure[!hbt](topskip=0pt, botskip=0pt, midskip=0pt)[width=6.5in]{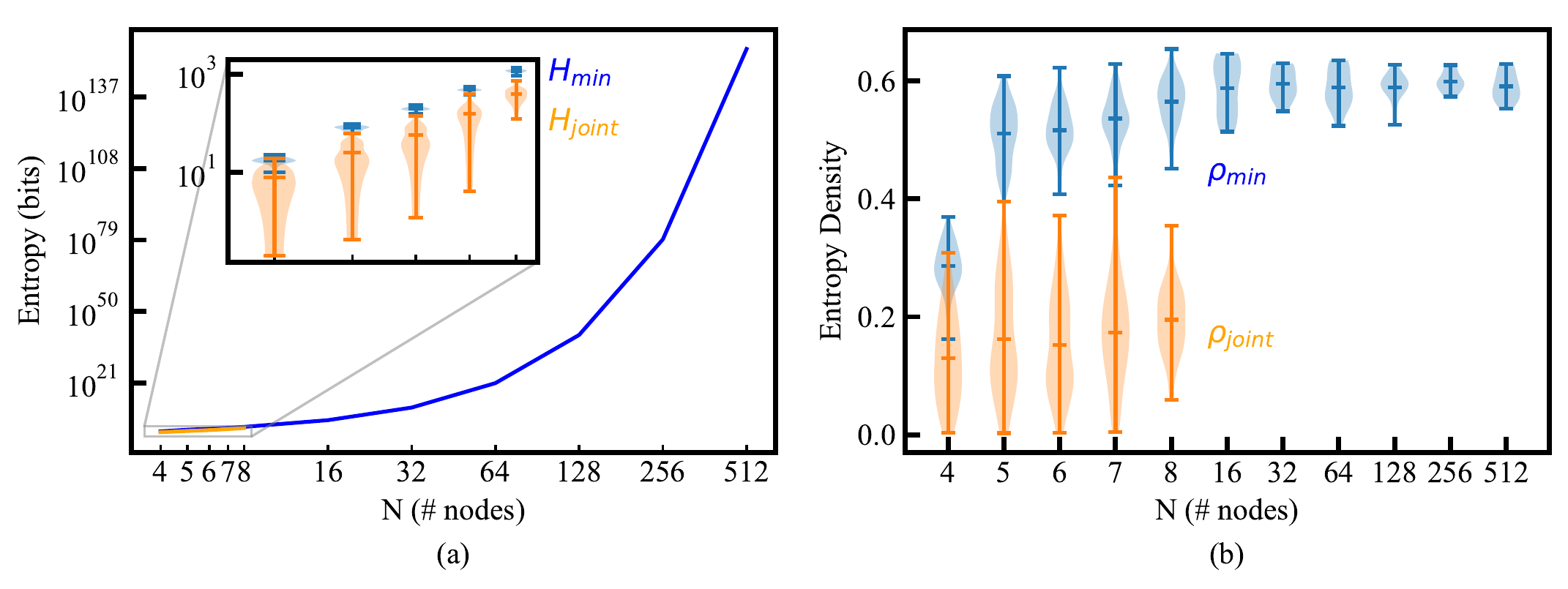}
{Entropy (a) measures and (b) densities of HBN-PUF classes as a function of network size $N$. Violin plots are over the distribution of classes, and solid lines indicate an average over classes.\label{fig:ENTROPY}}

\begin{table}[bth]
\centering
\caption{Entropies $H_{joint}\leq H_{min}$ and entropy densities $\rho_{i}\sim H_{i}/(N2^{N})$ for $N =4 - 512$. Only $H_{min}$ is estimated for $N>8$.}
\begin{tabular}{|c c c c c|}
\hline
 $N$ & $H_{min}$ & $\rho_{min}$ & $H_{joint}$ & $\rho_{joint}$ \\
 \hline\hline
    4 & $17.74 \pm 0.27$ & $0.32 $ & $8.04 \pm 0.59$ & $0.14 $ \\
    \hline
    5 & $80.69 \pm 0.89$ & $0.54 $ & $25.63 \pm 1.74$ & $0.17 $ \\
    \hline
    6 & $197.26 \pm 1.78$ & $0.53 $ & $58.10 \pm 3.70$ & $0.16 $ \\
    \hline
    7 & $479.40 \pm 3.88$ & $0.54 $ & $154.53 \pm 9.01$ & $0.18 $ \\
    \hline
    8 & $1155.88 \pm 9.29$ & $0.57 $ & $398.78 \pm 14.22$ & $0.20 $ \\
    \hline
    16 & $(6.17 \pm 0.12) \times10^{5}$ & $0.59$ & & \\
    \hline
    32 & $(8.18 \pm 0.08)\times10^{10}$ & $0.60$ & & \\
    \hline
    64 & $(6.95 \pm 0.09)\times10^{20}$ & $0.59$ & & \\
    \hline
    128 & $(2.56 \pm 0.02)\times10^{40}$ & $0.59$ & & \\
    \hline
    256 & $(1.77 \pm 0.01)\times10^{79}$ & $0.60$ & & \\
    \hline
    512 & $(4.06 \pm 0.00)\times10^{156}$ & $0.59$ & & \\
 \hline
\end{tabular}
\label{tab:ENTROPY}
\end{table}

Table \ref{tab:ENTROPY} records the entropy and entropy density, $\rho_{min}$ or $\rho_{joint}$, defined as the fraction of the observed entropy to the maximum possible entropy $N2^{N}$. We see that the entropy density for our median estimate $H_{min}$ hovers around 0.6, suggesting that the number of extractable bits is roughly $N2^{N}/2$ and hence that the min entropy scales super-exponentially with network size. Note however that there are theoretical bounds to the maximum entropy of PUFs and indeed any physical system, with arguments to be made that the entropy must be bounded polynomially by its size, such as the number of atoms \cite{STRONGPUF}. What our measurements show is that in the range $N=4-8$, for which entropy measures are calculated exactly over all possible CRPs, we observe super-exponential scaling with $N$. Outside this region, the entropy is computationally infeasible to calculate, and the reported values are extrapolations from limited measurements - which may not reflect the true entropy bounds of the system. 

The inset to Fig. \ref{fig:ENTROPY} illustrates the distribution of these entropy measures over 80 PUF classes for the exactly calculable network sizes $N=4-8$ (see Appendix \ref{app:EXP}). We observe that there is significant variation in the entropy estimates at very small PUF sizes, and that the joint entropy estimate in this region is approximately $15-20\%$ of full entropy. Note, however, that the joint entropy density increases and tightens as $N$ increases. We expect it to approach $\rho_{min}$ for larger networks.

We expect $\rho_{joint}$ to approach $\rho_{min}$ for two reasons.  Firstly, larger networks ($N>16$) consistently exhibit chaos, while small ABNs ($N\leq8$) may enter non-chaotic periodic regimes \cite{ROSINTHESIS} that induce pair-wise correlations.  Secondly, there exist certain challenge strings that are steady-state fixed points.  For the odd-input XOR functions used in this work, the all-zero and all-one challenge strings are fixed points; this can be seen since the output of the 3-XOR is zero or one if all its inputs are zero or one.  (In the case of an even number of inputs, the all-one challenge is not a fixed point.)  These trivial fixed points are filtered by our analysis, but there may exist other fixed points based on the details of the network wiring diagram that would need to be searched for via Boolean satisfiability algorithms which is not done in this work.  We expect the density of these fixed points to go to zero as $N\rightarrow\inf$, but a non-negligible fraction of the challenge space at the industrially-irrelevant network sizes shown in the inset may be steady-state fixed points that reduces the entropy. We see some evidence of this in the observed tightening of both entropy distributions with increasing $N$, and by the super-exponential growth of the extrapolated $H_{min}$ curve at larger sizes (see Appendix \ref{app:MINENT}).

Investigating these hypotheses and developing other means of estimating the entropy from limited samples for large networks is the subject of future work, as the exponential growth of the challenge space prevents full exploration even in principle.

\subsection{Machine Learning Attack with PUFmeter}
\label{subsec:AI}
PUFmeter \cite{PUFMETER} is a recently designed machine learning platform used to assess the security of a PUF. It attempts to learn the challenge-response behavior of a given PUF using probably-approximately-correct learning, and indicates whether a PUF's behavior can be learned and hence is susceptible to various attacks without actually performing specific attacks. The theory behind PUFmeter is based upon single-bit responses. For this reason, we use PUFmeter to assess the security of an individual bit of our responses to an attack, as well as the XOR of our entire response string. These results are presented in Table \ref{tab:AI}.

\begin{table}[!hbt]
\centering
\caption{PUFmeter machine-learning attack on an N=16 node HBN-PUF with responses taken after 6 pairs of inverter gates, using PUFmeter parameters $\delta=0.01$ and $\epsilon=0.05$ governing the probability thresholds for the analysis. Abbreviations Noise Upper Bound (UB), Average Sensitivity (AS), and Noise Sensitivity (NS). The result $\kappa=0$ indicates a failure of PUFmeter to model our PUF.}
\begin{tabular}{|c c c c c|}
\hline
 Response Bit & UB & AS & NS & $\kappa$ \\
 \hline\hline
 XOR & 0.468 & 0.298 & 0.249 & 0 \\ 
 \hline
 0th & 0.469 & 0.316 & 0.246 & 0 \\
 \hline
\end{tabular}
\label{tab:AI}
\end{table}
In Table \ref{tab:AI}, $\kappa$ is the minimum number of Boolean variables usable by PUFmeter to predict the response to a given challenge.  Because $\kappa=0$, PUFmeter is unable to model the behavior of the HBN-PUF. The noise upper bound, average sensitivity, and noise sensitivity are used to gauge the theoretical bounds for the types of attacks that are expected to be possible. From these results, PUFmeter indicates that an $N=16$ HBN-PUF may be susceptible to a Fourier-based attack. 

Summarizing, the observed super-exponential entropy scaling, the presence of chaotic nonlinear dynamics, and the failure of PUFmeter to model our PUF suggests that the behavior of the HBN-PUF may be resilient to machine learning attack. We have attempted machine learning attacks, including deep learning-based methods and model-based attacks, which have also failed and will be described in future publications. Further study is required to explicitly rule out any given attack, such as Fourier-based attacks and side-channel attacks. In such cases, instantiating multiple HBN-PUFs on the chip may obscure the power supply draw or the EM radiation emitted due to the chaotic transients of nearby networks.

\subsection{Cherry Picking and Temperature Variation}
\label{subsec:TEMP}
\Figure[!htb](topskip=0pt, botskip=0pt, midskip=0pt)[width=5.25in]{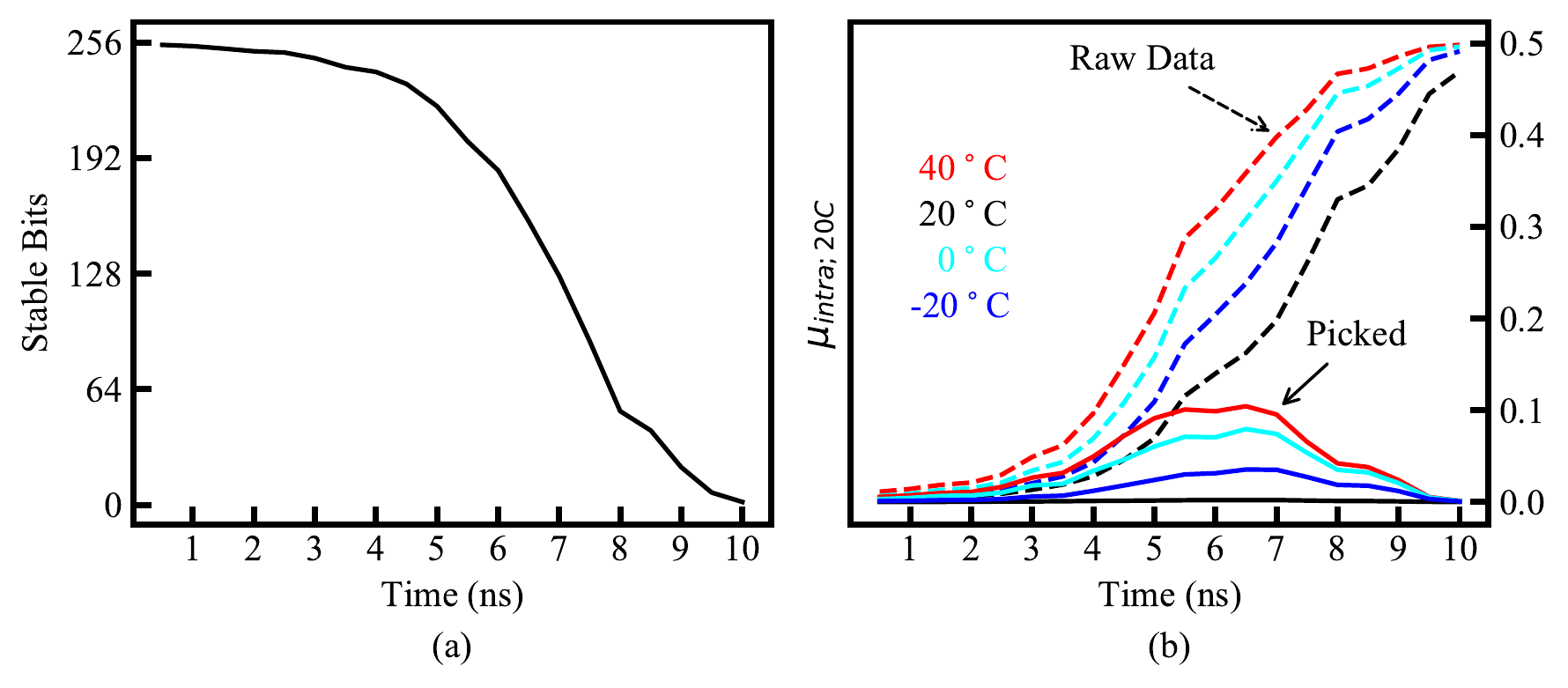}
{(a) The number of cherry picked stable bits vs. time for an $N=256$ network.  Stable bits are those that have less than 1\% error rate.  (b) $\mu_{intra}$ calculated with respect to a room temperature enrollment (20 $\degree$C) vs. measurement time for the same network when queried at different temperatures.  Dashed lines correspond to $\mu_{intra}$ calculated without cherry picking, and solid lines are with cherry picking. \label{fig:TEMP}}

A simple method of reducing errors is to mask out unreliable bits on a \textit{per challenge and per device} basis, an approach known as cherry picking \cite{CHERRY}.  That is, at enrollment, each PUF is queried multiple times (100 in this case) and any bits that vary are discarded; the bit mask used to discard bits is stored as helper data for reconstructing the PUF response at query time.   Fig. \ref{fig:TEMP}(a) shows the number of bits retained by this procedure (termed `stable bits') as a function of measurement time.  As can be seen, for measurement times up to about 7 ns, more than half of the 256 bits are stable at an error rate of less than 1\%.

We illustrate the usefulness of this cherry picking approach when querying the HBN-PUF at different temperatures, which is an important practical concern when comparing PUFs in different environmental conditions or over long operating times \cite{TEMPREF}.  A single $N=256$ HBN-PUF on a single chip was enrolled at room temperature (20 $\degree$C), and $\mu_{intra}$ was calculated with respect to this enrollment at three additional temperatures (-20 $\degree$C, 0 $\degree$C, and 40 $\degree$C, see Appendix \ref{app:TEMP}).  This $\mu_{intra;\mathrm{20\ }\degree\mathrm{C}}$ is plotted vs. measurement time in Fig. \ref{fig:TEMP}(b) in dashed lines, and compared to a control of a second collection at 20 $\degree$C (black).  We see that indeed there is an increased error rate compared to the control.  It is significant in the case of raw data; however, the cherry picking procedure (solid lines) does significantly reduce the error due to temperature variation.

The HBN-PUF has some degree of environmental stability due to the use of the delay line for triggering the capture of the network state.  Because the delay line is based on the same digital logic building blocks as the rest of the ABN, it is likely affected by temperature and voltage effects (\textit{e.g.}, changing rise, fall, and signal propagation times) in a way similar to the rest of the ABN.  Thus, if the entire network sped up or slowed down, the delay line would speed up or slow down in a commensurate way.  Contrast this with, \textit{e.g.}, an external, temperature-stabilized clock signal.  Early designs using a clock signal rather than delayed reset showed $\mu_{intra}$ close to 50\% for small temperature changes, but the delay line design is much more robust.

Strategies to reduce environmental variation, as well as experiments to test voltage sensitivity and aging effects, are future work for the HBN-PUF.  Referring to Fig. \ref{fig:TEMP}, we see that there is a trade-space between entropy/response (\textit{i.e.}, shorter measurement time corresponds to less entropy), error rate, bits/response, and effect temperature range that can be optimized over for specific applications.   This observation suggests that we can trade some of those bits for error correction ability to reduce errors to a level needed for key exchange because of the large number of bits available per response.  Moreover, the temperature effects \textit{do not} appreciably change the overall behavior of the PUF.  That is, there exists a $t_{opt}$ (that is constant for a PUF class over temperature) corresponding to $\Delta \mu \sim 0.35$ at any given temperature; it is changes to the specific bitstream, not differences in qualitative behavior, that drives these errors.  As a result, a temperature-aware enrollment protocol, in which the HBN-PUF is enrolled at multiple temperatures may be applicable \cite{MULTIPLETEMP}.

\section{Conclusions and Future Work}
\label{sec:CONC}
In summary, we present a novel HBN-PUF design that maps the challenge-response mechanism of the PUF onto the full state-space of a chaotic dynamical system (the HBN). The HBN-PUF represents an improvement in the state-of-the-art for strong PUFs  several ways. First, to our knowledge, the HBN-PUF is the only strong PUF proposal that produces multiple bits per response, thus reducing time, network, and storage resources for authentication and key exchange.  This will also likely frustrate machine learning attacks, as illustrated by our tests with PUFmeter, because the attacker will need to guess an $N$-dimensional Boolean vector instead of a one-dimensional one.  Second, the HBN-PUF is fast: response readout occurs in less than 10ns, which combined with the multiple bits per response, means that Gbps key generation rates are easily achievable.  Finally, the HBN-PUF is relatively insensitive to placement on the FPGA chip and resource usage scales linearly with the size of the PUF.  As a result, $N=1024$ or larger HBN-PUFs are easily realizable within resource constraints on modern low-end FPGAs (Cyclone V), but could produce upwards of $2^{1024}$ independent cryptographic keys at a rate of 100 Gbps.  This is fast enough so that, for instance, modern communications networks could be one-time pad encrypted with HBN-PUF output, but with such a large CRP space that it would take many lifetimes of the universe to exhaust the entropy.

The HBN-PUF has many attractive properties that suggest that it could be a true, machine-learning resistant and practical strong PUF.  However, there remain substantial questions to be addressed in future work.  The most obvious is further environmental testing and development of error mitigation strategies that are applicable to the HBN-PUF.  On a more theoretical level, the multiple bits per response stress existing entropy estimation methods and will require new techniques to more accurately lower-bound the actual extractable entropy. Moreover, we need to test and confirm the hypothesis that HBN-PUFs are in fact chaotic to prove the security properties of the HBN-PUF.  We have developed models of HBN-PUFs that can reproduce the behavior described here, which will appear in a follow up study, and we will use these models to execute model-based attacks to demonstrate machine learning resistance.  In addition to this theoretical work, a study of the effects of the network layout (\textit{e.g}, random vs. ring vs. other possible topologies) and detailed placement of the HBN-PUF elements in terms of the optimal measurement time and entropy per response will also appear in follow up work.

\appendix{}
\section{}

\subsection{Experimental Procedure}
\label{app:EXP}
The HBN-PUF is created by coding our design using the hardware description language Verilog (code in Appendix \ref{app:ABN}) using the Quartus CAD software, which compiles our code with automatic placement and routing chosen by its optimization procedure. We then program $N_{chips}=\Nchips$ separate DE10-Nano SOCs hosting Cyclone V 5CSEBA6U23I7 FPGAs with the same \textit{.sof} file.  This ensures each FPGA instantiates an identical copy of our PUF in both layout and design, meaning the only variations of instances within a PUF class are due to variations in the manufacturing of the FPGAs.    

For each network size $N$, we instantiate $N_{classes} = N_{graph}N_{loc}$ different HBN-PUF classes, where each class corresponds to a particular network topology randomly drawn from the set of possible regular graphs of degree 3 ($N_{graph}$ draws) and/or a particular location of the PUF on the chip ($N_{loc}$ PUFs per random graph). These draws are performed using custom Python scripts and the numpy.random module and written to the indicated positions in the Verilog file in Appendix \ref{app:ABN}.

In order to reduce the dependence on random seeds in the CAD's optimization procedures for the experiments presented here, we fix the locations of the nodes in the network to specific logic elements on the chip (which are randomly chosen from within a grid) but nothing about the HBN-PUF's behavior requires detailed control of node placement.  For $N\leq16$, $N_{loc}=16$; else, $N_{loc}=3$. For all sizes, $N_{graph} = 5$.  We create one \textit{.sof} file per random graph and place $N_{loc}$ PUFs at different locations (each with the same graph layout) on the chip in order to populate the distribution of HBN-PUFs. We find that the variation due to location is comparable to the variation due to graph layout, and so treat these on an equal footing; this yields a total of $N_{classes}=80$ different HBN-PUF classes for $N\leq16$ and $15$ HBN-PUF classes for $N>16$.

The Cyclone V chips that we use have an integrated hard processor running Linux.  We therefore use Altera's Avalon interface to make the PUF accessible to the Linux system and collect CRPs using custom C code that presents $N_{challenges}$ to each PUF via this interface to set the initial state of a given HBN. The HBN is held at a challenge for several $200$ MHz clock cycles due to synchronous controller logic and to stabilize the dynamics of the autonomous nodes. The network is then released and evolves for a short time during the transient phase, and the state is registered at a given delay time by choosing the length of the delay line via a multiplexer.  The response is transferred and and the PUF is reset to the same challenge.  The entire process is repeated $N_{repeats}=\Nrepeats$ times before moving to the next challenge, so that the total number of applied challenges to each HBN is equal to $N_{challenges} \times N_{repeats}$.  

Peculiar to the XOR function, there are two steady-state fixed points corresponding to when the network is all 0 or all 1. These fixed points are discarded from the challenge space as they have no entropy, however they can be used to identify `glitchy' PUF classes.  That is, since the HBN-PUF violates most commonly accepted design rules (in particular the guidance against large combinatorial loops), occasionally the Quartus software produces glitchy designs.  If a given PUF class does not produce all-ones or all-zeros as the response to an all-one or all-zero challenge, we discard the PUF class from consideration.  This occurs approximately ~10\% of the time.  All metrics are calculated using the valid challenges,  $N_{vc}=2^{N}-2$. For $N<16$, $N_{challenges}=N_{vc}$. For $N\geq16$, $N_{challenges}=1000$ unique and randomly selected valid challenges.  In all cases, $N_{repeats} = 100$.

These parameters are used for all experimental data collection unless otherwise noted.

\subsection{Formal Challenge-Response Definitions}
\label{app:CRP}
Let $P \in \mathbb{P}$ be a particular PUF instance $P$ belonging to the set of all PUF instances $\mathbb{P}$ of a particular PUF class. The response $\textbf{R}$ is a random variable $\textbf{R} : \mathbb{S}_{P} \rightarrow \{0,1\}^{N}$  mapping from the set of all possible physical states $\mathbb{S}_{P}$ of PUF instance $P$ to the set of all binary strings of length $N$, denoted $\{0,1\}^{N}$.  Specifically, the response takes as input a particular state $S_{P,C} \in \mathbb{S}_{P}$ of PUF instance $P$ resulting from challenge $\textbf{C}\in\{0,1\}^{N}$.

We characterize the reliability and uniqueness of $\mathbb{P}$ by studying the distributions of $\textbf{R}$ for various $P$ and $\textbf{C}$.  That is, we study how our design performs as a PUF by comparing responses from individual and different instances on a per-challenge basis using the metrics defined in the next appendix.

\subsection{Intra- and Inter-Device Statistics Definitions}
\label{app:STATS}
The degree to which two binary strings are different is given by the Hamming distance:
\begin{equation}
D(\textbf{A}, \textbf{B}) = \sum_{i=1}^N \textbf{A(i)} \oplus \textbf{B(i)},
\end{equation}
where $\textbf{A}$ and $\textbf{B}$ are the two binary strings to compare, of length $N$, $\textbf{A(i)}$ and $\textbf{B(i)}$ refer to the $i$-th bits of $\textbf{A}$ and $\textbf{B}$, respectively, and $\oplus$ is the XOR function.  For random strings, the Hamming distance is on average $N/2$.  Moreover, it is convenient to normalize the Hamming distance by $N$: $d(\textbf{A}, \textbf{B}) = D(\textbf{A}, \textbf{B})/N$.  For random strings $\textbf{A}$ and $\textbf{B}$, $d(\textbf{A}, \textbf{B}) = 1/2$.

Consider two different responses from the same challenge string $\textbf{C}_{c}$. These responses may result from applying the same challenge string to the same PUF instance (indexed by $p$) two different times (indexed by $r$ for repetition), $\textbf{R}_c^{p,r}$ and $\textbf{R}_c^{p,r'}$, or they may result from applying the challenge exactly once to two different PUF instances, $\textbf{R}_c^{p,r}$ and $\textbf{R}_c^{p',r}$.  Repeated application used to gauge reliability: a single PUF instance should ideally produce identical responses when presented with the same challenge (\textit{i.e.}, $d(\textbf{R}_i^{p,r}, \textbf{R}_i^{p,r'}) = 0$ for all $p, r$, and $r'$). Applying the same challenge to different PUF instances is used to gauge uniqueness: two different PUF instances should give responses to the same challenge which, when compared, appear random and uncorrelated.  In terms of Hamming distances, $d(\textbf{R}_i^{p,r}, \textbf{R}_i^{p',r}) \approx 1/2$ (although this does not capture correlations in bits).

For clarity we summarize these indices:
 \begin{itemize}
     \item $c \in [0,N_{challenges})$: Distinct challenge;
     \item $r,r' \in [0, N_{repeats})$: Separate applications of distinct challenge;
     \item $p,p' \in [0,N_{chips})$: Separate PUF instances.
 \end{itemize}
If we take each response to be an $N$-bit string, then the fraction of dissimilar bits between the two responses is denoted as

\begin{equation}
    \mathfrak{R}(c, p,r,r') = d(\textbf{R}_c^{p, r} , \textbf{R}_c^{p, r'}),
\end{equation}
\begin{equation}
    \mathfrak{U}(c, p,p',r) = d(\textbf{R}_c^{p, r} , \textbf{R}_c^{p', r}).  
\end{equation}

Above, $\mathfrak{R}$ (mnemonic `reliability') is the intra-device fractional Hamming distance between responses for the fixed PUF instance $p$ resulting from applications $r$ and $r'$ of challenge $c$. Likewise, $\mathfrak{U}$ (mnemonic `uniqueness') is the inter-device fractional Hamming distance between responses of PUF instances $p$ and $p'$ resulting from the fixed application $r$ of challenge $c$.

To obtain distributions of these distances on a per-challenge basis, we average over the pairwise combinations used to construct them, and then further average over the remaining indices to obtain mean measures of reliability $\mu_{intra}$ and uniqueness $\mu_{inter}$. Specifically, if we let $\langle \cdot \rangle_{a,b}$ indicate the average of a quantity over indices $a,b$, then

\begin{equation}
    \label{intraeq}
    \mathfrak{r}(c) =\langle \mathfrak{R}(c, p, r, r')\rangle _{r,r',p},
\end{equation}
\begin{equation}
    \label{intereq}
    \mathfrak{u}(c) =\langle \mathfrak{U}(c, p, p', r) \rangle _{p,p',r}.
\end{equation}

We record a time series of $N$-bit strings representing the time evolution of the network, so that the metrics introduced above exist at every measurement time. If we wish to measure the reliability on a per-chip basis, we simply do not average over $p$ in \eqref{intraeq}.

Fig. \ref{fig:MU} shows the histograms of \eqref{intraeq} and \eqref{intereq} at time $t_{opt}$. We further summarize the reliability and uniqueness as single numbers by averaging \eqref{intraeq} and \eqref{intereq} over challenges, \textit{i.e.},

\begin{equation}
    \mu_{intra} = \langle \mathfrak{r}(c) \rangle_{c},
\end{equation}
\begin{equation}
    \mu_{inter} = \langle \mathfrak{u}(c) \rangle_{c}.
\end{equation}

\subsection{Minimum Entropy}
\label{app:MINENT}
The min-entropy of a random variable $X$ is defined as
\begin{equation}
    H_{min}(X) = -\log(p_{max}(X)),
\end{equation}
where $p_{max}(X)$ is the probability of the most likely outcome. If $X = (x_{1}, x_{2}, ..., x_{n})$ is a vector of $n$ independent random variables, then the min-entropy is
\begin{equation}
    H_{min}=\sum_{i=1}^{n}-\log(p_{max}(x_{i})).\label{HMIN}
\end{equation}
In the case of a strong PUF with multiple challenges and a large response space, we need an ordering of the response bits in order to make sense of entropy calculations. A natural ordering is to define the response of the $i$-th node to the $j$-th challenge as $x_{jN+i}$, where the challenges are ordered lexicographically. This is illustrated in Table 3 for the simple case of $N = 3$. Here, there are only 6 challenges because we omit the all-0 and all-1 challenges as discussed in Appendix \ref{app:EXP}. 

\begin{table}[ht]
\centering
\caption{An illustration of response-bit ordering for $N = 3$, where there are $3\times6 = 18$ total bits.}
\begin{tabular}{|c c c c|}
\hline
 Challenge & Node 1 & Node 2 & Node 3 \\
 \hline\hline
 001 & $x_{1}$ & $x_{2}$ & $x_{3}$ \\ 
 \hline
 010 & $x_{4}$ & $x_{5}$ & $x_{6}$ \\
 \hline
 011 & $x_{7}$ & $x_{8}$ & $x_{9}$ \\
 \hline
 100 & $x_{10}$ & $x_{11}$ & $x_{12}$ \\
 \hline
 101 & $x_{13}$ & $x_{14}$ & $x_{15}$ \\  
 \hline
 110 & $x_{16}$ & $x_{17}$ & $x_{18}$ \\  
 \hline
\end{tabular}
\end{table}

Assuming independence of $x_{i}$, the min-entropy for the HBN-PUF can be readily calculated with \eqref{HMIN} from empirical estimates of $p_{max}(x_{i})$ \cite{SRAM, MINENT}. For each $x_{i}$, the estimate of $p_{max}(x_{i})$ is simply the observed frequency of 0 or 1, which ever is larger. To put the entropy calculations into context, we also present them as a fraction of the optimal case. If all of the $x_{i}$ were independent and completely unbiased, \textit{i.e.}, each $x_{i}$ were equally likely to be 0 or 1 (\textit{i.e.}, $p(x_i)=1/2$), then the min-entropy would be equal to  $N$ times the number of valid challenges $N_{vc}$. We therefore define the min-entropy density as
\begin{equation}
    \rho_{min}=H_{min}/(NN_{vc}).
\end{equation}
Due to the exponential scaling of the challenge space, we do not measure these values using all of the possible valid challenges for $N>8$. This is because of the computing time required in both calculating the entropy measures and obtaining the full CRP space. For $N>8$, we randomly choose challenges from a representative sample and multiply by the fraction of the unused space to obtain $H_{min}$. In the next appendix, we study the full challenge space for low $N$.

\subsection{Joint Entropy}
\label{app:JOINTENT}
In the previous appendix, we assume hat $x_{i}$ are independent, though this need not be the case. It is possible that some bits reveal information about others, reducing the entropy. Here we study these correlations between bit pairs, first by calculating the mutual information defined as
\begin{equation}
    I(x_{i},x_{j})=\sum_{x_{i},x_{j}}p(x_{i},x_{j})\log[\frac{p(x_{i},x_{j})}{p(x_{i})p(x_{j})}]
\end{equation}
between all pairs of $x_{i}$, $x_{j}$. Unlike min-entropy, the mutual information is difficult to calculate for higher $N$, so we will restrict our attention to $N = 4 - 8$ and use the full valid challenge space.

An adversary can use knowledge of any structure in the mutual information to more effectively guess response bits, thereby reducing the available entropy. In particular, the entropy is reduced to \cite{ENTROPYDEF}
\begin{equation}
    H_{joint}=H_{min}-\sum_{i=0}^{n-1}I(x_{i},x_{i+1}),
\end{equation}
where the ordering of the bits is such that the penalty is as large as possible. Calculating the ordering of the bits to maximize the joint information penalty is effectively a traveling salesman problem, which we solve approximately with a 2-opt algorithm \cite{KOPT}.

\subsection{Context-Tree Weighting Test}
\label{app:CTWENT}
In this appendix, we estimate the entropy through a string compression test. The results here should be understood as an upper-bound for the true entropy, especially for larger $N$. In particular, we consider the context tree weighting (CTW) algorithm \cite{CTW1}. 

The CTW algorithm takes a binary string called the context and forms an ensemble of models that predict subsequent bits in the string. It then losslessly compresses subsequent strings into a codeword using the prediction model. The size of the codeword is defined as the number of additional bits required to encode the PUF instance’s challenge-response behavior. If the context contains information about a subsequent string, then the codeword will be of reduced size.

In the case of PUFs, the codeword length approaches the true entropy of the generating source in the limit of unbounded tree depth \cite{CTW2}. However, the required memory scales exponentially with tree depth, so it is not computationally feasible to consider an arbitrarily deep tree in the CTW algorithm. Instead, we vary the tree depth up to 20 to optimize the compression.

We perform a CTW compression as follows:
\begin{itemize}
    \item We collect data for $N = 4 - 8$ HBN-PUFs with $N_{repeats} = 1$.
    \item We concatenate the resulting measurements for all but one PUF instances into a 1D
string of length $(N_{chips}-1)N_{vc}N$ to be used as context.
    \item We apply the CTW algorithm to compress the measurements from the last PUF
with the context, using various tree depths to optimize the result.
    \item We repeat steps 2-3, omitting measurements from a different PUF instance, until all
PUFs have been compressed.
\end{itemize}
The final entropy estimate is the average codeword length from all of the compression tests described above. If the behavior of the $N_{chips} - 1$ PUF instance can be used to predict the behavior of the unseen instance, then the PUFs do not have full entropy.

\subsection{Temperature Variation}
\label{app:TEMP}

We calculate at each temperature the deviation of an HBN-PUF with respect to itself at 20  $\degree$C, a quantity which we denote $\mu_{intra;\mathrm{20}\ ^\degree\mathrm{C}}$. This measure is equivalent to considering an individual chip as consisting of different instances - one for each temperature. It is calculated at each temperature by comparing responses to those generated at 20  $^\degree$C, then averaging over all challenges. These plots are presented in Fig. \ref{fig:TEMP} as a function of $t$, the number of inverter gates after which the response is registered. Each curve is a separate temperature.

\subsection{Hardware Description Language Code}
\label{app:ABN}
This Verilog code is used for synthesizing the HBN in Fig. \ref{fig:HBNPUF}.
\begin{lstlisting}[language=Verilog]
//This module corresponds to the node zoom-in of Fig. 2.
module Node(reset, challenge, in1, in2, in3, out);

   input reset;
   input challenge;
   input in1;
   input in2;
   input in3;
   output out;

   wire node_clocked;
   wire node_asynchronous;

   assign node_asynchronous = in1 ^ in2 ^ in3;
   assign node_clocked = reset ? challenge : node_asynchronous;
   assign out = node_clocked;

endmodule


module HBN(

    clk,
	reset,
    challenge,
	delay_address,
    response, 
	ready

);

parameter N = 256;    // HBN Size
parameter INDEX = 0; // Index on the chip (0..2 or 0..15 depending on N)            
parameter MEASUREMENT_DELAY = 20;

localparam DELAY_ADDRESS_BITS = $clog2(MEASUREMENT_DELAY);

input  	        clk;     	  
input           reset;       
input [N-1:0] 	challenge;   
input [DELAY_ADDRESS_BITS:0] 	delay_address;
output [N-1:0] 	response;    
output reg 		ready;


wire [N-1:0] ring_state;  
wire [N-1:0] ring [3]; 
wire         reset_buf /*synthesis keep*/;
wire         delayed_reset /*synthesis keep*/;


assign reset_buf = reset;

// create N nodes
genvar i;
generate
for (i=0; i<N; i=i+1) begin : generate_ring
   Node n (
	
	   .reset(reset_buf),
		.reset_delay(delayed_reset),
		.challenge(challenge[i]),
		.in1(ring[0][i]),
		.in2(ring[1][i]),
		.in3(ring[2][i]),
		.out(ring_state[i]),
		.response(response[i])
	
	);
end
endgenerate


always @(posedge clk) begin
	if( reset ) ready <= 0;
	else ready <= ~delayed_reset;
end

AddressableDelayLine #(MEASUREMENT_DELAY) DLM (reset, delay_address, delayed_reset);

/// AUTO-GENERATED CODE TO DEFINE THE WIRING DIAGRAM
assign ring[0][0] = ring_state[5];      // 0th input of node 0
assign ring[1][0] = ring_state[37];     // 1st input of node 0
assign ring[2][0] = ring_state[131];    // 2nd input of node 0

/// ...and so-on...
\end{lstlisting}
This Verilog code is used for synthesizing the tapped-delay line in Fig. \ref{fig:HBNPUF}.
\begin{lstlisting}[language=Verilog]
// This is the tapped delay line in Fig. 1.  Reconstructing the time series requires resetting the PUF and incrementing the delay_address
module AddressableDelayLine(
   in,
   delay_address,
   out
);

parameter N = 5;    // # of pairs of inverters.
localparam DELAY_ADDRESS_BITS = $clog2(N);

input 	             			in;           
input[DELAY_ADDRESS_BITS:0] 	delay_address;
output    	          			out;          

wire [2*N-1:0] delay /*synthesis keep*/;

assign delay[0] = in;
assign out = delay[2*delay_address-1];

genvar i;
    generate for (i=0; i<2*N-1; i=i+1) begin : generate_delays
       assign delay[i+1] = ~delay[i];
    end
    endgenerate

endmodule
\end{lstlisting}

\bibliographystyle{IEEEtran}
\bibliography{IEEEabrv,References}

\begin{thebibliography}{10}
\providecommand{\url}[1]{#1}
\csname url@samestyle\endcsname
\providecommand{\newblock}{\relax}
\providecommand{\bibinfo}[2]{#2}
\providecommand{\BIBentrySTDinterwordspacing}{\spaceskip=0pt\relax}
\providecommand{\BIBentryALTinterwordstretchfactor}{4}
\providecommand{\BIBentryALTinterwordspacing}{\spaceskip=\fontdimen2\font plus
\BIBentryALTinterwordstretchfactor\fontdimen3\font minus
  \fontdimen4\font\relax}
\providecommand{\BIBforeignlanguage}[2]{{%
\expandafter\ifx\csname l@#1\endcsname\relax
\typeout{** WARNING: IEEEtran.bst: No hyphenation pattern has been}%
\typeout{** loaded for the language `#1'. Using the pattern for}%
\typeout{** the default language instead.}%
\else
\language=\csname l@#1\endcsname
\fi
#2}}
\providecommand{\BIBdecl}{\relax}
\BIBdecl

\bibitem{PUFDEF}
\BIBentryALTinterwordspacing
R.~Maes, \emph{Physically Unclonable Functions: Concept and
  Constructions}.\hskip 1em plus 0.5em minus 0.4em\relax Springer Berlin
  Heidelberg, 2013, ch.~2, pp. 25--48. [Online]. Available:
  \url{http://link.springer.com/10.1007/978-3-642-41395-7_2}
\BIBentrySTDinterwordspacing

\bibitem{TAXONOMY}
\BIBentryALTinterwordspacing
T.~McGrath, I.~E. Bagci, Z.~M. Wang, U.~Roedig, and R.~J. Young, ``A {PUF}
  taxonomy,'' \emph{Applied Physics Reviews}, vol.~6, no.~1, p. 011303, Mar
  2019. [Online]. Available:
  \url{http://aip.scitation.org/doi/10.1063/1.5079407}
\BIBentrySTDinterwordspacing

\bibitem{CRYPTOKEY}
\BIBentryALTinterwordspacing
J.~Zhang, T.~Q. Duong, A.~Marshall, and R.~Woods, ``Key generation from
  wireless channels: A review,'' \emph{IEEE Access}, vol.~4, pp. 614--626,
  2016. [Online]. Available: \url{http://ieeexplore.ieee.org/document/7393435/}
\BIBentrySTDinterwordspacing

\bibitem{SRAM}
\BIBentryALTinterwordspacing
D.~Holcomb, W.~Burleson, and K.~Fu, ``Power-up {SRAM} state as an identifying
  fingerprint and source of true random numbers,'' \emph{IEEE Transactions on
  Computers}, vol.~58, no.~9, pp. 1198--1210, Sep 2009. [Online]. Available:
  \url{http://ieeexplore.ieee.org/document/4674345/}
\BIBentrySTDinterwordspacing

\bibitem{EARLYPUF}
\BIBentryALTinterwordspacing
R.~Pappu, ``Physical one-way functions,'' \emph{Science}, vol. 297, no. 5589,
  pp. 2026--2030, Sep 2002. [Online]. Available:
  \url{https://www.sciencemag.org/lookup/doi/10.1126/science.1074376}
\BIBentrySTDinterwordspacing

\bibitem{FPGAPUF}
\BIBentryALTinterwordspacing
S.~S. Kumar, J.~Guajardo, R.~Maes, G.-J. Schrijen, and P.~Tuyls, ``Extended
  abstract: The butterfly {PUF} protecting {IP} on every {FPGA}.''\hskip 1em
  plus 0.5em minus 0.4em\relax IEEE, Jun 2008, pp. 67--70. [Online]. Available:
  \url{http://ieeexplore.ieee.org/document/4559053/}
\BIBentrySTDinterwordspacing

\bibitem{STRONGPUF}
\BIBentryALTinterwordspacing
U.~Rührmair, H.~Busch, and S.~Katzenbeisser, \emph{Strong {PUF}s: Models,
  Constructions, and Security Proofs}, ser. Information Security and
  Cryptography.\hskip 1em plus 0.5em minus 0.4em\relax Springer Berlin
  Heidelberg, 2010, ch. chapter 4, pp. 79--96. [Online]. Available:
  \url{http://link.springer.com/10.1007/978-3-642-14452-3_4}
\BIBentrySTDinterwordspacing

\bibitem{CHAOSPUF}
\BIBentryALTinterwordspacing
K.~Gołofit and P.~Wieczorek, ``Chaos-based physical unclonable functions,''
  \emph{Applied Sciences}, vol.~9, no.~5, p. 991, Mar 2019. [Online].
  Available: \url{https://www.mdpi.com/2076-3417/9/5/991}
\BIBentrySTDinterwordspacing

\bibitem{CHAOSFPGA}
\BIBentryALTinterwordspacing
T.~Tuncer, ``The implementation of chaos-based {PUF} designs in field
  programmable gate array,'' \emph{Nonlinear Dynamics}, vol.~86, no.~2, pp.
  975--986, Oct 2016. [Online]. Available:
  \url{http://link.springer.com/10.1007/s11071-016-2938-3}
\BIBentrySTDinterwordspacing

\bibitem{CHAOSENHANCE}
\BIBentryALTinterwordspacing
L.~Chen, ``A framework to enhance security of physically unclonable functions
  using chaotic circuits,'' \emph{Physics Letters A}, vol. 382, no.~18, pp.
  1195--1201, May 2018. [Online]. Available:
  \url{https://linkinghub.elsevier.com/retrieve/pii/S037596011830255X}
\BIBentrySTDinterwordspacing

\bibitem{EDSBOOK}
E.~Ott, \emph{Fractal Basin Boundaries}.\hskip 1em plus 0.5em minus 0.4em\relax
  Cambridge University Press, Aug 1993, ch. 5.1, pp. 152--157.

\bibitem{HBNTRNG}
\BIBentryALTinterwordspacing
D.~P. Rosin, D.~Rontani, and D.~J. Gauthier, ``Ultrafast physical generation of
  random numbers using hybrid boolean networks,'' \emph{Physical Review E},
  vol.~87, no.~4, Apr 2013. [Online]. Available:
  \url{https://link.aps.org/doi/10.1103/PhysRevE.87.040902}
\BIBentrySTDinterwordspacing

\bibitem{ROSINTHESIS}
\BIBentryALTinterwordspacing
D.~P. Rosin, \emph{Autonomous Boolean Networks on Electronic Chips}, ser.
  Springer Theses.\hskip 1em plus 0.5em minus 0.4em\relax Springer
  International Publishing, 2015, ch. chapter 3, pp. 25--33. [Online].
  Available: \url{http://link.springer.com/10.1007/978-3-319-13578-6_3}
\BIBentrySTDinterwordspacing

\bibitem{BCHAOS1}
\BIBentryALTinterwordspacing
R.~Zhang, H.~L.~D. de~S.Cavalcante, Z.~Gao, D.~J. Gauthier, J.~E.~S. Socolar,
  M.~M. Adams, and D.~P. Lathrop, ``Boolean chaos,'' \emph{Physical Review E},
  vol.~80, no.~4, Oct 2009. [Online]. Available:
  \url{https://link.aps.org/doi/10.1103/PhysRevE.80.045202}
\BIBentrySTDinterwordspacing

\bibitem{GHIL}
\BIBentryALTinterwordspacing
M.~Ghil and A.~Mullhaupt, ``Boolean delay equations. ii. periodic and aperiodic
  solutions,'' \emph{Journal of Statistical Physics}, vol.~41, no. 1-2, pp.
  125--173, Oct 1985. [Online]. Available:
  \url{http://link.springer.com/10.1007/BF01020607}
\BIBentrySTDinterwordspacing

\bibitem{BCHAOS2}
\BIBentryALTinterwordspacing
H.~L. D. d.~S. Cavalcante, D.~J. Gauthier, J.~E.~S. Socolar, and R.~Zhang, ``On
  the origin of chaos in autonomous boolean networks,'' \emph{Philosophical
  Transactions of the Royal Society A Mathematical Physical and Engineering
  Sciences}, vol. 368, no. 1911, pp. 495--513, Jan 2010. [Online]. Available:
  \url{https://royalsocietypublishing.org/doi/10.1098/rsta.2009.0235}
\BIBentrySTDinterwordspacing

\bibitem{CHERRY}
\BIBentryALTinterwordspacing
M.~Hiller, M.-D. Yu, and G.~Sigl, ``Cherry-picking reliable {PUF} bits with
  differential sequence coding,'' \emph{IEEE Transactions on Information
  Forensics and Security}, vol.~11, no.~9, pp. 2065--2076, Sep 2016. [Online].
  Available: \url{http://ieeexplore.ieee.org/document/7480444/}
\BIBentrySTDinterwordspacing

\bibitem{ENTROPYDEF}
\BIBentryALTinterwordspacing
R.~Maes, \emph{{PUF}-Based Key Generation}.\hskip 1em plus 0.5em minus
  0.4em\relax Springer Berlin Heidelberg, 2013, ch. chapter 6, pp. 143--168.
  [Online]. Available:
  \url{http://link.springer.com/10.1007/978-3-642-41395-7_6}
\BIBentrySTDinterwordspacing

\bibitem{PUFMETER}
\BIBentryALTinterwordspacing
F.~Ganji, D.~Forte, and J.-P. Seifert, ``{PUF}meter a property testing tool for
  assessing the robustness of physically unclonable functions to machine
  learning attacks,'' \emph{IEEE Access}, vol.~7, pp. 122\,513--122\,521, 2019.
  [Online]. Available: \url{https://ieeexplore.ieee.org/document/8819883/}
\BIBentrySTDinterwordspacing

\bibitem{TEMPREF}
\BIBentryALTinterwordspacing
S.~K. Mathew, S.~K. Satpathy, M.~A. Anders, H.~Kaul, S.~K. Hsu, A.~Agarwal,
  G.~K. Chen, R.~J. Parker, R.~K. Krishnamurthy, and V.~De, ``16.2 a 0.19pj/b
  pvt-variation-tolerant hybrid physically unclonable function circuit for 100%
  stable secure key generation in 22nm cmos.''\hskip 1em plus 0.5em minus
  0.4em\relax IEEE, Feb 2014, pp. 278--279. [Online]. Available:
  \url{http://ieeexplore.ieee.org/document/6757433/}
\BIBentrySTDinterwordspacing

\bibitem{MULTIPLETEMP}
\BIBentryALTinterwordspacing
Y.~Gao, Y.~Su, L.~Xu, and D.~C. Ranasinghe, ``Lightweight (reverse) fuzzy
  extractor with multiple reference {PUF} responses,'' \emph{IEEE Transactions
  on Information Forensics and Security}, vol.~14, no.~7, pp. 1887--1901, Jul
  2019. [Online]. Available:
  \url{https://ieeexplore.ieee.org/document/8574914/}
\BIBentrySTDinterwordspacing

\bibitem{MINENT}
\BIBentryALTinterwordspacing
P.~Simons, E.~van~der Sluis, and V.~van~der Leest, ``Buskeeper {PUF}s, a
  promising alternative to d flip-flop {PUF}s.''\hskip 1em plus 0.5em minus
  0.4em\relax IEEE, Jun 2012, pp. 7--12. [Online]. Available:
  \url{http://ieeexplore.ieee.org/document/6224311/}
\BIBentrySTDinterwordspacing

\bibitem{KOPT}
\BIBentryALTinterwordspacing
B.~Chandra, H.~Karloff, and C.~Tovey, ``New results on the old {K}-opt
  algorithm for the traveling salesman problem,'' \emph{SIAM Journal on
  Computing}, vol.~28, no.~6, pp. 1998--2029, Jan 1999. [Online]. Available:
  \url{http://epubs.siam.org/doi/10.1137/S0097539793251244}
\BIBentrySTDinterwordspacing

\bibitem{CTW1}
F.~M.~J. Willems, Y.~M. Shtarkov, and T.~J. Tjalkens, ``The context tree
  weighting method: Basic properties,'' \emph{IEEE Trans. Inf. Theory},
  vol.~41, no.~3, pp. 653--664, May 1995.

\bibitem{CTW2}
\BIBentryALTinterwordspacing
T.~Ignatenko, G.-j. Schrijen, B.~Skoric, P.~Tuyls, and F.~Willems, ``Estimating
  the secrecy-rate of physical unclonable functions with the context-tree
  weighting method.''\hskip 1em plus 0.5em minus 0.4em\relax IEEE, Jul 2006,
  pp. 499--503. [Online]. Available:
  \url{http://ieeexplore.ieee.org/document/4036011/}
\BIBentrySTDinterwordspacing

\end{thebibliography}

\begin{IEEEbiography}[{\includegraphics[width=1in,height=1.25in,clip,keepaspectratio]{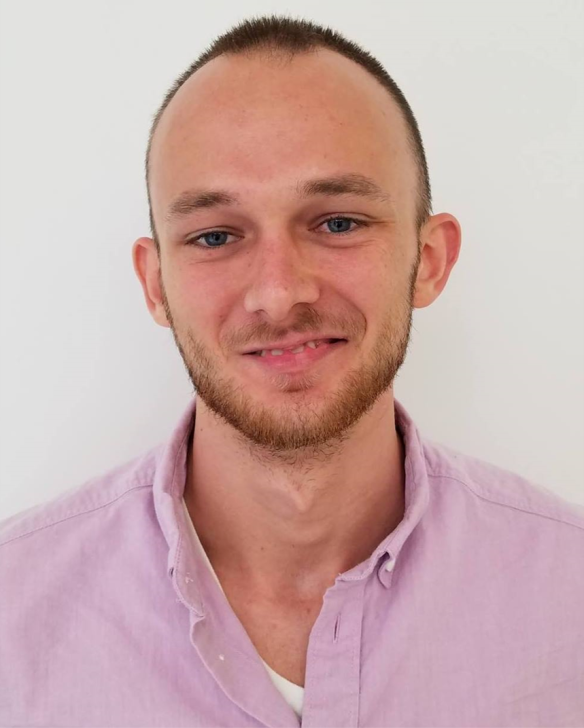}}]{Noeloikeau F. Charlot} 
received the B.S. degrees in physics and biological engineering from the University of Hawai’i at Manoa, Honolulu, HI, USA in 2017, and an M.S. degree in physics from the Ohio State University, Columbus, OH, USA in 2020, where he is currently pursuing a Ph.D. degree in physics. 

Before his work as a research assistant in the QuantInfo lab and CYAN cybersecurity collaboration at OSU, he interned in the Gravitational Astrophysics lab at NASA Goddard and the Dark Matter Detection lab at UH Manoa. Noelo’s industrial experience includes bioreactor design and photonics at ProtaCulture, LLC. His research interests include network science, artificial intelligence, ultrafast electronics, and quantum gravity. Noelo is a prior McNair Scholar.

\end{IEEEbiography}

\begin{IEEEbiography}[{\includegraphics[width=1in,height=1.25in,clip,keepaspectratio]{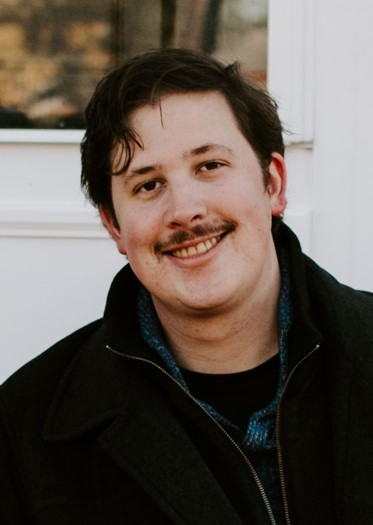}}]{Daniel Canaday}
was born in Columbus, OH in 1991. He received the B.S. degree in physics and mathematics from Ohio State University, Columbus, OH, USA in 2014, the M.S. degree in physics from Ohio State University, Columbus, OH, USA in 2017, and the Ph.D. degree in physics from Ohio State University, Columbus, OH, USA in 2019.

He is currently a scientist at Potomac Research, LLC, Alexandria, VA. His research is concerned with applied reservoir computing and the application of physical neural networks to cryptography.
\end{IEEEbiography}

\begin{IEEEbiography}[{\includegraphics[width=1in,height=1.25in,clip,keepaspectratio]{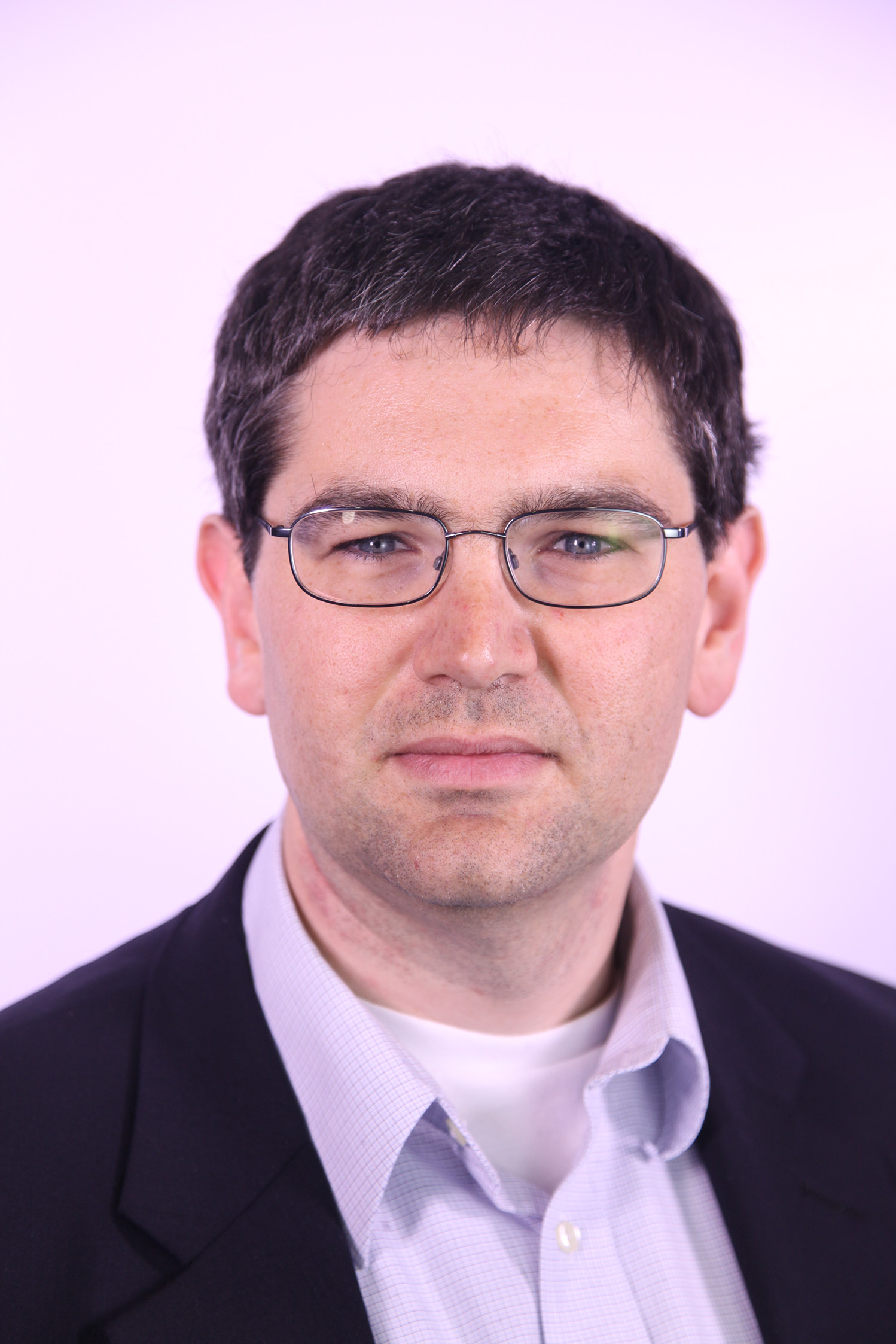}}]{Andrew Pomerance}
Andrew Pomerance  was born in Washington, DC, USA, in 1980. He received the B.S. and M.S. degrees in Electrical and Computer Engineering from Carnegie Mellon University, Pittsburgh, PA, in 2002, and the Ph.D. in Physics from the University of Maryland, College Park, MD, in 2009. 

From 2009 to 2013, he was with Raytheon Applied Signal Technology, Tyson's Corner, VA, USA. He is currently the president of Potomac Research, LLC, Alexandria, VA, USA.  His research is concerned with nonlinear dynamics with applications to machine learning and cryptography.
\end{IEEEbiography}

\begin{IEEEbiography}[{\includegraphics[width=1in,height=1.25in,clip,keepaspectratio]{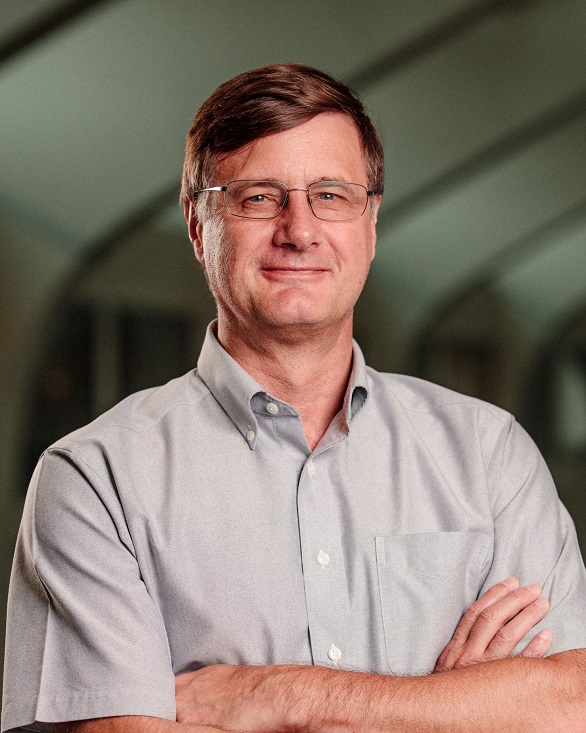}}]{Daniel J. Gauthier}
Daniel J. Gauthier is a Professor of Physics and Electrical and Computer Engineering at The Ohio State University. He received the B.S., M.S., and Ph.D. degrees from the University of Rochester, Rochester, NY, in 1982, 1983, and 1989, respectively. His Ph.D. research on “Instabilities and chaos of laser beams propagating through nonlinear optical media” was supervised by Prof. R.W. Boyd and supported in part through a University Research Initiative Fellowship. From 1989 to 1991, he developed the first CW two-photon optical laser as a Post-Doctoral Research Associate under the mentorship of Prof. T.W. Mossberg at the University of Oregon. In 1991, he joined the faculty of Duke University, Durham, NC, as an Assistant Professor of Physics and was named a Young Investigator of the U.S. Army Research Office in 1992 and the National Science Foundation in 1993.  He was the Robert C. Richardson Professor of Physics at Duke from 2011- 2015, chair of the Duke Physics Department from 2005 – 2011, interim chair in spring 2015, and was a founding member of the Duke Fitzpatrick Institute for Photonics. He moved to The Ohio State University in 2016. His research interests include: reservoir computing, synchronization and control of the dynamics of complex networks in electronic and optical systems, quantum communication, and nonlinear quantum optics. Prof. Gauthier is a Fellow of the Optical Society of America and the American Physical Society.
\end{IEEEbiography}

\EOD

\end{document}